\documentstyle[preprint,tighten,eqsecnum,aps,floats,psfig,epsfig,amssymb,prb]{revtex}

\begin{document}
\draft
\title{
Spin-density-wave order in cuprates
}
\author{Martino De Prato,${}^{1,2}$ 
       Andrea Pelissetto,${}^3$  Ettore Vicari${}^4$ }
\address{$^1$
Max-Planck Institut f\"ur Metallforschung, Heisenbergstrasse 3, D-70569
Stuttgart, Germany}
\address{$^2$
Inst. f\"ur Theoretische und Angewandte Physik, Universit\"at 
Stuttgart, \\ Pfaffenwaldring 57, D-70569
Stuttgart, Germany}
\address{$^3$ Dipartimento di Fisica dell'Universit\`a di Roma 
``La Sapienza" and INFN,\\ 
P.le Aldo Moro 2, I-00185 Roma, Italy}
\address{$^4$
Dipartimento di Fisica dell'Universit\`a di Pisa 
and INFN, \\ 
Largo Pontecorvo 2, I-56127 Pisa, Italy
}
\address{
{\bf e-mail: 
{\tt Deprato@mf.mpg.de},
{\tt Andrea.Pelissetto@roma1.infn.it},
{\tt Ettore.Vicari@df.unipi.it}
}}

\date{\today}

\maketitle

\begin{abstract}
We study the nature of the two-dimensional quantum critical point separating
two phases with and without long-range spin-density-wave order, which has
been recently observed in cuprate superconductors.  We consider the
Landau-Ginzburg-Wilson Hamiltonian associated with the 
spin-density critical modes,
perform a mean-field analysis of the phase diagram, and study the
corresponding renormalization-group flow in two different perturbative
schemes at five and six loops, respectively. The analysis supports the
existence of a stable fixed point in the full theory whose basin of
attraction includes systems with collinear spin-density-wave order, as
observed in experiments.  The stable fixed point is characterized by an
enlarged ${\rm O(4)}\otimes{\rm O(3)}$ symmetry. The continuous transition
observed in experiments is expected to belong to this universality class.
The corresponding critical exponents are $\nu = 0.9(2)$ and $\eta =
0.15(10)$.
\end{abstract}



\section{Introduction}\label{sec1}

In the last few decades several aspects of cuprate superconductors (SCs)
have been studied and many efforts have been spent to understand the
unique and complex phase diagram exhibited by this class of materials;
see, e.g., Ref.~\CITE{Sachdev-03}.  Superconductivity in cuprates
appears to be due to a mechanism analogous to the BCS one in ordinary
superconductors. However, superconductivity is only one of the
characteristic features of these materials. There are many other new
properties that require more complex mechanisms and can be understood
only if the interplay between BCS and additional order parameters is
considered. For instance, at $T\approx 0$,
$\mbox{La}_{2-\delta}\mbox{Sr}_\delta\mbox{CuO}_4$ at very low doping
$\delta$ is an insulator with long-range magnetic order. Increasing
$\delta$, at $\delta\approx0.055$ an insulator-superconductor first-order
transition takes place, giving rise to a superconducting state in
which spins are still magnetically ordered
\cite{expt}. At $\delta \approx 0.14$
another phase transition occurs, and, for $\delta\gtrsim 0.14$, the material
shows no magnetic order---it is paramagnetic---but is still
superconducting.  Neutron-scattering
experiments\cite{AMHMK-97} suggested that this transition is continuous.
Moreover, in the ordered phase $\delta \lesssim 0.14$, they
revealed the presence of collinearly polarized spin-density waves (SDWs) 
with wavevectors
\begin{eqnarray}
&& {\bf K}_1 = \frac{2\pi}{a}\,\left(\frac{1}{2} -\theta,\frac{1}{2} \right), 
\qquad\qquad
   {\bf K}_2=   \frac{2\pi}{a}\,\left(\frac{1}{2} ,\frac{1}{2} -\theta\right), 
\label{eq_kappas} 
\end{eqnarray}
where $\theta$ is a function of the doping concentration and $a$ is
the lattice spacing.  The wave vectors ${\bf K}_{i}$ are
two-dimensional since cuprates are supposed to be made of weakly
interacting planes and thus behave approximately as two-dimensional systems.
Following Ref.~\CITE{ZDS-02}, we assume that superconductivity is not 
relevant at the transition which is instead driven by the 
interaction among the SDW degrees of freedom. Since $T\approx 0$
one should take into account the quantum nature of the system. Quantum
phase transitions can be studied by introducing a supplementary
dimension parametrized by an imaginary time variable $\tau$. 
The relevant order parameter is the spin field which is parametrized as
\begin{equation}
S_i({\bf r},\tau)= {\rm Re}[e^{i {\bf K}_1 \cdot {\bf r}}\,
\Phi_{1i}({\bf r},\tau)+e^{i {\bf K}_2 \cdot {\bf r}}\,
\Phi_{2i}({\bf r},\tau)],
\end{equation}
where $\Phi_{ai}$ are complex amplitudes.  There are two interesting
limiting cases.  The first one is when the order
parameter can be written as $\Phi_{a}({\bf r},\tau)=e^{i\alpha_a}~{\bf
n}_a,$ which corresponds to collinearly polarized SDWs.  The second one is
when $\Phi_{a}({\bf r},\tau)~=~{\bf n}_{a,1}+i \ {\bf n}_{a,2}$, with
${\bf n}_{a,1}\cdot {\bf n}_{a,2}=0$ and $|{\bf n}_{a,1}|=|{\bf n}_{a,2}|$, 
which corresponds to circularly polarized SDWs.
In cuprates experiments indicate that the ground state shows a collinear 
behavior.\cite{Yee-etal-99}

The standard strategy for writing down an effective Hamiltonian for a
given physical system consists in considering all polynomials of the
order parameter of order less than or equal to four that are
compatible with the expected symmetries.  In the SDW-SC--to--SC phase
transition the order parameter is the complex 
field $\Phi_{ai}({\bf r},\tau)$, with $a=1,2$ and $i=1,2,3$. The
corresponding symmetries are the following:
(i) SO(3) spin rotations:
$\Phi_{ai}~\rightarrow~ O_{ij}\,\Phi_{aj}$;
(ii) Translational symmetry of the spin waves:
$\Phi_{ai} \rightarrow e^{i \alpha_a}\,\Phi_{ai}$;
(iii) Spatial inversion:
$\Phi_{ai} \rightarrow  \Phi_{ai}^*$;
(iv) Interchange of the $\hat{1}$ and $\hat{2}$ axes:
$\Phi_{1i} \leftrightarrow \Phi_{2i}$ and $x \leftrightarrow y$.
The most general Hamiltonian with these symmetries is\cite{ZDS-02}
\begin{eqnarray}
\mathcal{H} &=&\int d^2 r\, d\tau \left\{ \vert\partial_\tau
\Phi_{1}\vert^2 +v_1^2\vert\partial_x \Phi_{1}\vert^2
+v_2^2\vert\partial_y \Phi_{1}\vert^2 +\vert\partial_\tau
\Phi_{2}\vert^2+ \right.\nonumber\\ && \left.+v_2^2\vert\partial_x
\Phi_{2}\vert^2 +v_1^2\vert\partial_y \Phi_{2}\vert^2
+r(\vert\Phi_{1}\vert^2+\vert\Phi_{2}\vert^2)+\right.\nonumber\\
&& \left.+\frac{u_{1,0}}{2}(\vert\Phi_{1}\vert^4+\vert\Phi_{2}\vert^4)
+\frac{u_{2,0}}{2}(\vert\Phi_{1}^2\vert^2+\vert\Phi_{2}^2\vert^2)+
\right.\nonumber\\
&& \left.
  +w_{1,0}\vert\Phi_{1}\vert^2 \vert\Phi_{2}\vert^2
  +w_{2,0}\vert\Phi_{1} \cdot \Phi_{2}\vert^2 
  +w_{3,0}\vert\Phi_{1}^* \cdot \Phi_{2}\vert^2 \right\},
\label{lgwhor} 
\end{eqnarray}
where $v_1$ and $v_2$ are parameters called SDW velocities. Terms such
as $\Phi^*_{a}\cdot \partial_\tau \Phi_{a}$ are forbidden by spatial
inversion symmetry and terms like $i \Phi^*_{a}\cdot \partial_x
\Phi_{a}$, even if permitted by all symmetries, can be eliminated 
by redefining the fields as 
$\Phi_{a} \to e^{i {\mathbf q}_a \cdot {\mathbf r}} \Phi_{a}$.
Hamiltonian (\ref{lgwhor}) admits several different ground states 
depending on the values of the parameters. They are classified in 
App.~\ref{AppA}. In particular, there is the possibility that 
both fields correspond to collinearly polarized SDWs as observed 
in experiments:
$\Phi_1 = e^{i\alpha_1} {\mathbf n}_1$ and 
$\Phi_2 = e^{i\alpha_2} {\mathbf n}_2$, where the vectors 
${\mathbf n}_1$ and ${\mathbf n}_2$
satisfy either ${\mathbf n}_1 = {\mathbf n}_2$
or ${\mathbf n}_1 \cdot {\mathbf n}_2 = 0$. 

In this paper we investigate the nature of the fixed points (FPs) of 
the renormalization-group (RG) flow of the effective Hamiltonian (\ref{lgwhor}).
If a stable FP
exists and its attraction domain includes systems with collinearly polarized 
SDWs, then the SDW-SC--to--SC transition may be continuous.
Otherwise, it must be of first order.  In our study, we 
consider only the case $v_1 = v_2$ that simplifies the analysis and allows 
us to perform a high-order perturbative analysis. Therefore, we consider the 
theory
\begin{eqnarray}
\mathcal{H} &=&\int d^d x \left\{ 
\sum_\mu^d \left(\vert\partial_\mu \Phi_{1}\vert^2 +
                 \vert\partial_\mu \Phi_{2}\vert^2 \right)
+r(\vert\Phi_{1}\vert^2+\vert\Phi_{2}\vert^2 )
+\frac{u_{1,0}}{2}(\vert\Phi_{1}\vert^4+\vert\Phi_{2}\vert^4)
\right.\nonumber\\
&&\left. +\frac{u_{2,0}}{2}(\vert\Phi_{1}^2\vert^2+\vert\Phi_{2}^2\vert^2)
   +w_{1,0}\vert\Phi_{1}\vert^2 \vert\Phi_{2}\vert^2
   +w_{2,0}\vert\Phi_{1}\cdot \Phi_{2}\vert^2 
   +w_{3,0}\vert\Phi_{1}^* \cdot \Phi_{2}\vert^2 
   \vphantom{\sum_\mu^d} \right\},
\label{lgwh} 
\end{eqnarray}
where the field
$\Phi_{ai}$ is a complex $2\times N$ matrix, $a=1,2$, $i=1,\ldots,N$.
The physically relevant case is $N=3$.

We first perform a standard analysis close to four dimensions,\cite{WF-72} 
computing the RG functions in powers of $\epsilon \equiv 4 - d$.
A one-loop analysis indicates that a stable FP exists only
for $N\gtrsim 42.8$. Apparently, this result casts doubts on the existence of a 
stable FP in three dimensions. 
However, in three dimensions there may exist FPs that are absent for $\epsilon
\ll 1$. This is indeed what happens in the Ginzburg-Landau model of
superconductors, in which a complex scalar field couples to a gauge
field\cite{superc} and in O(2)$\otimes$O($n$) symmetric 
models.\cite{CPPV-04,DPV-04} Thus, a more careful investigation of 
the RG flow in three dimensions calls for strictly three-dimensional 
perturbative schemes. For this purpose 
we consider two field-theoretical 
perturbative approaches: the minimal-subtraction scheme without
$\epsilon$ expansion\cite{SD-89} (in the following we will indicate
it as $3d$-$\overline{\rm MS}$ scheme) and the massive zero-momentum
(MZM) renormalization scheme.\cite{Parisi-80}
The use of two different schemes is crucial, since
the comparison of the corresponding results provides a nontrivial check
on the reliability of our conclusions. In the
$3d$-$\overline{\rm MS}$ scheme one considers the massless (critical)
theory in dimensional regularization,\cite{tHV-72} 
determines the RG functions from the divergences
appearing in the perturbative expansion of the correlation functions,
and finally sets $\epsilon\equiv 4-d=1$ without expanding in powers of
$\epsilon$ (this scheme therefore differs from the standard $\epsilon$
expansion\cite{WF-72}). In the MZM scheme one considers instead the
three-dimensional massive theory in the disordered
(high-temperature) phase. We compute the $\beta$
functions to five loops in the $3d$-$\overline{\rm MS}$
scheme and to six loops in the MZM scheme. 
We use a symbolic manipulation program that generates the diagrams 
(approximately one thousand at six loops) and
computes their symmetry and group factors, and the compilation of
Feynman integrals of Refs.~\CITE{NMB-77,KS-01}. 
The series are available on request. The perturbative 
expansions are then resummed using the known large-order behavior. 

The perturbative analysis of the RG flow in the full theory is 
not sufficiently stable to provide reliable results. Therefore,
we have focused on the 
stability of the FPs that occur in specific submodels of Hamiltonian 
(\ref{lgwh}). The analysis of the perturbative series indicates the 
stability of the ${\rm O(4)}\otimes{\rm O(3)}$ collinear FP
that occurs in the model with $w_{1,0} = u_{1,0} - u_{2,0}$ and 
$w_{2,0} = w_{3,0} = u_{2,0} < 0$. 
Moreover, its basin of attraction includes systems with 
collinear SDWs. Therefore, we expect the continuous transition observed 
experimentally in cuprates to belong to this universality class.
This implies an effective enlargement of the symmetry at the transition
point. The corresponding critical exponents would be 
\begin{equation}
\nu = 0.9(2), \qquad\qquad \eta = 0.15(10).
\end{equation}

The paper is organized as follows. In Sec.~\ref{sec2} we discuss the 
possible ordered phases that occur in model (\ref{lgwhor}) in the mean-field 
approximation. Details are given in App.~\ref{AppA}.
In Sec.~\ref{4drgflow} we discuss the FP structure 
close to four dimensions in the standard $\epsilon$ expansion. 
Sec.~\ref{substa} contains the main results of this work.
We consider three different submodels (Sec.~\ref{submodels})
and then investigate the stability properties of the FPs occuring in 
each of them (Sections \ref{stabo2on}, \ref{stabo4on}, and 
\ref{stabmn}). Conclusions are presented in Sec.~\ref{conclusions}.
In App.~\ref{appo4on} and \ref{appmn} we give some technical details.

\section{Mean-field analysis} \label{sec2}

The phase diagram of Hamiltonian (\ref{lgwhor}) can be studied in the 
mean-field approximation. Due to the large number of couplings the analysis is 
quite complex. We have limited our considerations to the case $N\le 3$. 
We summarize here the results that are derived in App.~\ref{AppA}. For 
$r > 0$ the system is disordered and $\Phi_1 = \Phi_2 = 0$. For $r = 0$
a continuous phase transition occurs followed by a magnetized phase 
with $r < 0$. The nature of the ordered phase depends on the values of the 
quartic parameters. The analysis reported in App.~\ref{AppA} shows that 
there are seven possibilities:
\begin{itemize}
\item[(1)] $\Phi_1$ is a collinear SDW ($\Phi_1 = e^{i\alpha_1} {\mathbf n}$, 
$\mathbf{n}$ real) while $\Phi_2 = 0$.
\item[(2)] $\Phi_1$ is a circularly polarized SDW 
($\Phi_1 = e^{i\alpha_1} ({\mathbf n}_1 + i {\mathbf n}_2)$,
${\mathbf n}_1$ and ${\mathbf n}_2$ real, 
$|{\mathbf n}_1| = |{\mathbf n}_2|$,
${\mathbf n}_1 \cdot {\mathbf n}_2 = 0$) while $\Phi_2 = 0$.
\item[(3)] $\Phi_1$ and $\Phi_2$ correspond to collinear SDWs
with the same axis and amplitude:
$\Phi_1 = e^{i\alpha_1} {\mathbf n}$, 
$\Phi_2 = e^{i\alpha_2} {\mathbf n}$, 
${\mathbf n}$ real.
\item[(4)] $\Phi_1$ and $\Phi_2$ correspond to collinear SDWs
with orthogonal axes and same amplitude:
$\Phi_1 = e^{i\alpha_1} {\mathbf n}_1$, 
$\Phi_2 = e^{i\alpha_2} {\mathbf n}_2$, 
${\mathbf n}_1, {\mathbf n}_2$ real, 
${\mathbf n}_1 \cdot {\mathbf n}_2 = 0$,
$|{\mathbf n}_1| = |{\mathbf n}_2|$.
\item[(5)] $\Phi_1$ and $\Phi_2$ are circularly polarized SDWs
with the same rotation plane and amplitude:
$\Phi_1 = e^{i\alpha_1} ({\mathbf n}_1 + i {\mathbf n}_2)$,
$\Phi_2 = e^{i\alpha_2} ({\mathbf n}_1 + i {\mathbf n}_2)$,
${\mathbf n}_1$ and ${\mathbf n}_2$ real, 
$|{\mathbf n}_1| = |{\mathbf n}_2|$,
${\mathbf n}_1 \cdot {\mathbf n}_2 = 0$.
\item[(6)] $\Phi_1$ is a collinear SDW and $\Phi_2$ is a circularly polarized 
SDW.
The rotation plane of $\Phi_2$ is orthogonal to the axis of  $\Phi_1$.
Explicitly:
$\Phi_1 = e^{i\alpha_1} {\mathbf n}_1$,
$\Phi_2 = e^{i\alpha_2} ({\mathbf n}_2 + i {\mathbf n}_3)$,
${\mathbf n}_i$ real,  
$|{\mathbf n}_2| = |{\mathbf n}_3|$,
${\mathbf n}_i \cdot {\mathbf n}_j = 0$.
\item[(7)] $\Phi_1$ and $\Phi_2$ are elliptically polarized SDWs
with different rotation planes but with the same amplitude,
$|\Phi_1| = |\Phi_2|$.
\end{itemize}
For cuprates the relevant solutions are (3) and (4). Necessary conditions
to obtain (3) are $w_{2,0} + w_{3,0} < 0$ and 
\begin{equation}
  w_{1,0} + w_{2,0} + w_{3,0} - u_{1,0} < u_{2,0} < \hbox{Min}\, 
  [u_{1,0} - w_{1,0} - w_{2,0} - w_{3,0}, -w_{2,0}, -w_{3,0}],
\label{cond21}
\end{equation}
while (4) requires $w_{2,0} + w_{3,0} > 0$ and 
\begin{equation}
  w_{1,0} - u_{1,0} < u_{2,0} < \hbox{Min}\, 
  [u_{1,0} - w_{1,0} , w_{2,0}, w_{3,0}].
\label{cond22}
\end{equation}
These conditions are not sufficient, since for some values of the 
parameters satisfying Eqs.~(\ref{cond21}) or (\ref{cond22}) 
the ordered phase is given
by solutions (6) or (7). Note that the sign of $u_{2,0}$ is not the 
relevant parameter that selects the collinear SDWs among all possible
solutions. 

It is interesting to note that the mean-field solution predicts either 
$\Phi_1 \| \Phi_2$ or $\Phi_1 \bot \Phi_2$ in the case of collinear SDWs.
This result is easy to understand. If both fields correspond to 
collinear SDWs, then one can take $\Phi_1$ and $\Phi_2$ real. In this
case the only term of the Hamiltonian that contains a scalar product of the 
two fields is $(w_{2,0} + w_{3,0}) (\Phi_1 \cdot \Phi_2)^2$ that forces the 
two fields to be either parallel or orthogonal, depending on the sign of 
$w_{2,0} + w_{3,0}$. 
Note that this also holds if we add additional higher-order 
terms to the Hamiltonian, as long as the transition is continuous. 
Indeed, for a continuous transition $\Phi_a\to 0$ at the transition 
($\Phi_a = 0$ in the 
disordered phase) and thus higher-order terms do not play any role. 
On the other hand, this relation may not be valid if the transition is 
of first order. Also the coupling to the charge-density waves (CDWs) that are 
present in cuprates\cite{ZKE-98,ZDS-02} 
does not change this conclusion, since they
couple to the scalars $\Phi_a^2$, $|\Phi_a|^2$.

Solutions (3) and (4) also satisfy $|\Phi_1| = |\Phi_2|$. 
This property does not 
necessarily hold if we take into account the CDWs 
(see Refs.~\CITE{ZKE-98,ZDS-02} for an extensive discussion). 
Indeed, let $\phi_1$ and $\phi_2$ be 
the complex amplitudes of the CDWs coupled respectively to 
$\Phi_1^2$ and $\Phi_2^2$. In the absence of the CDW-SDW coupling, for some values 
of the CDW Hamiltonian parameters, the ordered solution corresponds to 
$|\phi_1| \not=0$, $\phi_2 = 0$. If now the CDW-SDW coupling is included, 
one may obtain a ground state with $|\phi_1| \not= |\phi_2| \not= 0$ and 
$|\Phi_1| \not= |\Phi_2| \not= 0$.

\section{RG flow close to four dimensions}
\label{4drgflow}

The RG flow close to four dimensions can be investigated perturbatively
in $\epsilon\equiv 4-d$.  In the minimal-subtraction ($\overline{\rm MS}$)
the one-loop $\beta$ functions are:
\begin{eqnarray}
&&
\beta_{u_1}=
- \epsilon u_1+(N+4)u_1^2+4u_1 u_2+4u_2^2+N 
w_1^2+w_2^2+w_3^2+2w_1w_2+2w_1w_3,
\nonumber 
\\
&&
\beta_{u_2}=-\epsilon u_2+6u_1u_2+N u_2^2+2 w_2 w_3,
\nonumber \\
&&
\beta_{w_1}=-\epsilon w_1+
    2 w_1^2+w_2^2+w_3^2+2(N+1)u_1 w_1+4 u_2 w_1+2u_1 w_2+2u_1w_3,
\nonumber \\
&&
\beta_{w_2}=-\epsilon w_2+N w_2^2+2 u_1 w_2+4u_2 w_3+4w_1 w_2+2 w_2 w_3,
\nonumber \\
&&
\beta_{w_3}=-\epsilon w_3 + N w_3^2+2u_1w_3+4 u_2 w_2+4w_1w_3+2w_2w_3,
\label{betaonel}
\end{eqnarray}
where $u_i,\,w_i$ are the renormalized quartic couplings corresponding
to the quartic Hamiltonian parameters $u_{i,0}$, $w_{i,0}$. 
They are normalized so that, at tree level, $g = g_0 \mu^{-\epsilon}/A_d$,
where $g$ and $g_0$ label the renormalized and Hamiltonian parameters
respectively and $A_d \equiv 2^{d-1} \pi^{d/2} \Gamma(d/2)$.
The FPs of the RG flow are the common zeroes of the
$\beta$ functions. 
For $N=3$ there are 4 FPs while for $N=2$ there are 7 FPs: they are all
unstable.  Only for $N\gtrsim 42.8$ does a
stable FP exist. It has $u_2 = w_2 = w_3$ (for $N\to \infty$ 
we obtain $u_1 = u_2 = w_2 = w_3 = \epsilon/N$, $w_1 = 0$),
so that at the FP the symmetry becomes ${\rm O(4)}\otimes{\rm O(}N{\rm )}$.
This FP is the chiral FP that occurs in 
${\rm O(}M{\rm )}\otimes{\rm O(}N{\rm )}$  in the large-$N$ 
limit.\cite{PRV-01b}

In order to determine the behavior in three dimensions, one should 
extend the computation to higher order in $\epsilon$ and determine
the function $N_c(\epsilon) = 42.8 + O(\epsilon)$ such that the chiral FP point 
identified above exists for $N > N_c(\epsilon)$ and is no longer 
present for smaller values of $N$. 
We have not pursued this approach for several reasons. 
First, the analogous five-loop computation that was performed 
in the ${\rm O}(N)\otimes{\rm O}(2)$ model\cite{Kawamura-98,PRV-01b,CP-04} 
was not able to explain the correct physics of these models for 
$N=2,3$ (see Sec.~II.D in Ref.~\CITE{CPPV-04}). 
Moreover, this calculation is only concerned with the stable FP that 
is present for $\epsilon = 0$ (in the present case the chiral
${\rm O}(4)\otimes{\rm O}(N)$ FP), while in $d = 3$ the stable FP may be 
different,
an unstable or even a new FP. The analysis that will be presented in the 
next Section favors this last possibility.

\section{Submodels and their stability}
\label{substa}

The three-dimensional properties of the RG flow are determined by its
FPs.  Some of them can be identified by considering particular cases
in which some of the quartic parameters vanish.  The corresponding FPs
are also FPs of the general theory.  In this section, we identify some
of them, and then determine their stability with respect to the
complete theory.

\subsection{Some particular cases}
\label{submodels}

For particular values of the couplings
Hamiltonian (\ref{lgwh}) reduces to that of simpler models. 
Three cases have already been extensively studied in the 
literature:\cite{PV-rev}
\begin{itemize}
\item[\mbox{(1)}] 
For $w_{1,0}=w_{2,0}=w_{3,0}=0$ there is no interaction between
the two SDWs and Hamiltonian (\ref{lgwh}) reduces to that of 
two identical decoupled O(2)$\otimes{\rm O}(N)$-symmetric models.  
The general ${\rm O}(m)\otimes{\rm O}(n)$-symmetric
model is defined by the Hamiltonian density\cite{Kawamura-98,PV-rev}
\begin{eqnarray}
  && {1\over2}
\sum_{ai} \left[ \sum_\mu (\partial_\mu \phi_{ai})^2 + r \phi_{ai}^2 
      \right]   
+ {g_{1,0}\over 4!}  \Bigl( \sum_{ai} \phi_{ai}^2\Bigr)^2 
\nonumber \\
&& \qquad\qquad
+ {g_{2,0}\over 4!}  \left[ \sum_{i,j} 
   \Bigl( \sum_a \phi_{ai} \phi_{aj}\Bigr)^2 - 
   \Bigl(\sum_{ai} \phi_{ai}^2 \Bigr)^2 \right],
\label{Hch}
\end{eqnarray}
where $\phi_{ai}$ is a real $n\times m$ matrix field
($a=1,\ldots,n$ and $i=1,\ldots,m$). 
Hamiltonian (\ref{Hch}) is obtained from Eq.~(\ref{lgwh}) by setting
$\Phi_{ai}= \phi_{1i}^{(a)}+ i \phi_{2i}^{(a)}$ and
\begin{equation}
u_{1,0}=g_{1,0}/3 - g_{2,0}/6, \quad 
u_{2,0}=g_{2,0}/6, \quad w_{1,0}=w_{2,0}=w_{3,0}=0\; .
\label{corre}
\end{equation}
The properties of ${\rm O}(2)\otimes{\rm O}(N)$ models are reviewed in
Refs.~\CITE{CPPV-04,Kawamura-98,PV-rev,DMT-03}.  In three dimensions
perturbative calculations within the MZM scheme\cite{PRV-01,CPS-02}
and within the $3d$-$\overline{\rm MS}$ scheme\cite{CPPV-04} indicate
the presence of a stable chiral FP with attraction domain in the
region $g_{2,0}>0$ for all values of $N$ (only for $N=6$ the evidence
is less clear, since the MZM analysis does not apparently support
it). For $N=2$, these conclusions have been recently confirmed by a
Monte Carlo simulation.\cite{CPPV-04}  A stable collinear FP for
$g_2<0$ exists for $N\le 4$.\cite{DPV-04,CPV-05}  Apart from the
collinear FP for $N=2$, these FPs do not exist close to four dimensions.
For $N=2$ the collinear FP is equivalent to an XY FP and corresponds to 
$g_1^* = g^*_{XY}$, $g_2^* = -g^*_{XY}$,
where $g^*_{XY}$ is the FP value of the renormalized coupling in the 
O(2) $\phi^4$ model.

\item[\mbox{(2)}] 
For $w_{1,0}=u_{1,0}-u_{2,0}$ and
$w_{2,0}=w_{3,0}=u_{2,0}$, Hamiltonian (\ref{lgwh}) reduces to 
(\ref{Hch}) with $m=4$ and $n=N$.  The correspondence is given by 
\begin{eqnarray}
\Phi_{1i} = {\phi_{1i}+ i \phi_{2i}\over \sqrt{2}},
\qquad 
\Phi_{2i} = {\phi_{3i}+ i \phi_{4i}\over \sqrt{2}},
\label{o4oncorr}
\end{eqnarray}
where $\phi_{ei}$ is a $4\times N$ matrix, and 
\begin{eqnarray}
g_{1,0}=3(u_{1,0}+u_{2,0}), \quad g_{2,0} = 6 u_{2,0} \ .
\label{o4oncorr2} 
\end{eqnarray}
We have already discussed the FPs of the ${\rm O}(4)\otimes{\rm O}(2)$
theory.  The ${\rm O}(4)\otimes{\rm O}(3)$ theory does not present stable FPs
for $g_2>0$.\cite{Parruccini-03} Analyses of the available
six-loop series in the MZM scheme and five-loop series in the 3-$d$
$\overline{\rm MS}$ scheme indicate the presence of a stable collinear FP
for $g_2<0$.\cite{foot2} This FP does not exist close to four dimensions.

\item[\mbox{(3)}]
For $u_{2,0}=w_{2,0}=w_{3,0}=0$ we obtain the $mn$ model with $n=2$ and $m=2N$.
The so-called $mn$ model is defined by the Hamiltonian 
density\cite{Aharony-76,PV-rev}
\begin{eqnarray}
  {1\over2}
\sum_{ai} \left[ \sum_\mu (\partial_\mu \phi_{ai})^2 + r \phi_{ai}^2 
      \right]   
+ {g_{1,0}\over 4!}  \Bigl( \sum_{ai} \phi_{ai}^2\Bigr)^2 
+ {g_{2,0}\over 4!}   \sum_{aij} \phi_{ai}^2 \phi^2_{aj},
\label{HMN}
\end{eqnarray}
where $\phi_{ai}$ is a real $n\times m$ matrix, i.e., $a=1,\ldots,n$
and $i=1,\ldots,m$. The correspondence is obtained by setting 
\begin{equation}
\Phi_{ai} = (\phi_{ai} + i \phi_{a,i+N})/\sqrt{2}, \qquad
g_{1,0}=3w_{1,0}, \qquad g_{2,0} = 3(u_{1,0}-w_{1,0}).  
\label{corr-mn}
\end{equation}
A stable FP is the O$(m)$ FP with $g_1 = 0$ and $g_2 = g_{{\rm O}(m)}^*$, where
$g_{{\rm O}(m)}^*$ is the FP value of the renormalized coupling in the 
$O(m)$-symmetric vector
model. In three dimensions the analysis of five- and six-loop 
series\cite{PV-05}  indicates the presence of a second stable FP 
with $g_2 < 0$ for $n = 2$ and $m=2$, 3, and 4.  
\end{itemize}

Beside these three models, there are two other submodels for which no results
are available:
\begin{itemize}
\item[(a)] For $w_{2,0} = w_{3,0} = 0$ 
we obtain two chiral models coupled by an 
energy-energy term.  Note that in this model the RG flow does not cross
the planes $u_{2} = 0$ and $w_{1} = 0$. 
\item[(b)] For $w_{2,0} = w_{3,0} = w_0$ we obtain a model with an additional
U(1) symmetry: $\Phi_1 \to \Phi_1^*$, $\Phi_2 \to \Phi_2^*$.
In this model the RG flow does not cross the plane $w = 0$.
\end{itemize}

Finally, note an additional symmetry of Hamiltonian (\ref{lgwh}). It 
is invariant under $\Phi_1 \to \Phi_1^*$, $\Phi_2 \to \Phi_2$, 
$w_{2,0} \to w_{3,0}$ and $w_{3,0} \to w_{2,0}$, while the other couplings
are unchanged. This implies that the RG flow in the space of renormalized 
couplings does not cross the plane $w_2 = w_3$
and that, for any FP with $w_2 > w_3$ there is an equivalent one 
with $w_2 < w_3$. In particular, we can limit our considerations to 
$w_2 \ge  w_3$.

In order to study the RG flow of the theory one can start by discussing
the stability in the full theory of the FPs of the models (1), (2), and (3) 
discussed above.

For $N=2$ and $N=3$, the only cases we consider, model (1) has two FPs:
\begin{itemize}
\item[(1a)] the chiral FP, in which $g_1 = g^*_{1,\rm ch}$
            and $g_2 = g^*_{2,\rm ch}$; correspondingly
            $u_1^* = g^*_{1,\rm ch}/3 - g^*_{2,\rm ch}/6$, 
            $u_2^* = g^*_{2,\rm ch}/6>0$, 
            $w_1^*=w_2^*=w_3^*=0$;
\item[(1b)] the collinear FP, in which $g_1 = g^*_{1,\rm cl}$
            and $g_2 = g^*_{2,\rm cl}$; correspondingly
            $u_1^* = g^*_{1,\rm cl}/3 - g^*_{2,\rm cl}/6$, 
            $u_2^* = g^*_{2,\rm cl}/6<0$, 
            $w_1^*=w_2^*=w_3^*=0$.
\end{itemize}
Here $g_{i,\rm ch}$ and $g_{i,\rm cl}$ are the chiral and collinear FPs
of the O(2)$\otimes$O($N$) theory. The analogous FPs are present in model (2):
\begin{itemize}
\item[(2a)] the chiral FP, in which $g_1 = g^*_{1,\rm ch}$
            and $g_2 = g^*_{2,\rm ch}$; correspondingly
            $u_1^* = g^*_{1,\rm ch}/3 - g^*_{2,\rm ch}/6$, 
            $u_2^* = g^*_{2,\rm ch}/6<0$, 
            $w_1^* = u_1^* - u_2^*$, 
            $w_2^* = w_3^* = u_2^*$;
            It does not exist for $N = 3$. This is the FP 
            that is relevant for $N>N_c(\epsilon)\approx 42.8 + O(\epsilon)$ 
            close to four dimensions;
\item[(2b)] the collinear FP, in which $g_1 = g^*_{1,\rm cl}$
            and $g_2 = g^*_{2,\rm cl}$; correspondingly
            $u_1^* = g^*_{1,\rm cl}/3 - g^*_{2,\rm cl}/6$, 
            $u_2^* = g^*_{2,\rm cl}/6<0$, 
            $w_1^* = u_1^* - u_2^*$, 
            $w_2^* = w_3^* = u_2^*$.
            It exists for both $N=2$ and $N=3$.
\end{itemize}
Here $g_{i,\rm ch}$ and $g_{i,\rm cl}$ are the chiral and collinear FPs
of the O(4)$\otimes$O($N$) theory. Finally, the $mn$ theory gives two 
FPs:
\begin{itemize}
\item[(3a)] the O(2$N$) FP. This is unstable in the full theory,
being already unstable in model (1);
\item[(3b)] the $mn$ FP $g_1 = g_{1,mn}$, $g_2 = g_{2,mn}$; correspondingly
            $u_1^* = g^*_{1,mn}/3 + g^*_{2,mn}/3$, $w_1^* = g^*_{1,mn}/3$,
            $u_2^*=w_2^*=w_3^*=0$.  It exists only for $N=2$.
\end{itemize}

In the following we study the stability of these FPs in the complete
theory (\ref{lgwh}). For this purpose, using the $\beta$ functions of the 
general theory we have computed the stability matrices of the FPs at 
six and five loops respectively 
in the MZM and $3d$-$\overline{\rm MS}$ schemes.
The perturbative series have been resummed by using the 
conformal-mapping method described, e.g., in Ref.~\CITE{ZJbook}.
For a FP belonging to a submodel,
the large-order behavior needed for the conformal-mapping summation 
is the same as that characterizing all series of that
submodel. For all submodels we consider, the
large-order behavior is already known.\cite{PRV-01,Parruccini-03,PV-00}

\subsection{Stability of the decoupled ${\rm O}(2)\otimes{\rm O}(N)$ fixed points}
\label{stabo2on}

We want to establish the stability properties of the decoupled 
${\rm O}(2)\otimes{\rm O}(N)$ FPs (1a) and (1b) in the complete theory
(\ref{lgwh}). For this purpose we need the RG dimensions of
the operators present in Hamiltonian
(\ref{lgwh}) that break the symmetry of model (1), i.e., of the operators
associated with the quartic couplings
$w_i$. It is useful to rewrite them as
\begin{eqnarray}
&&w_{1,0}\vert\Phi_{1}\vert^2 \vert\Phi_{2}\vert^2
+w_{2,0}\vert\Phi_{1} \cdot \Phi_{2}\vert^2 
+w_{3,0}\vert\Phi_{1}^*\cdot \Phi_{2}\vert^2 
= W_{00} P_{00} + W_{11} P_{11} + W_{02} P_{02} ,
\label{pert}
\end{eqnarray}
where 
\begin{equation}
W_{00}= w_{1,0}+\frac{1}{N}(w_{2,0}+w_{3,0}), \quad
W_{11}= -w_{2,0}+w_{3,0}, \quad
W_{02}= w_{2,0}+w_{3,0}, 
\end{equation}
\begin{equation}
P_{00}\equiv O_{00}^{(1)} O_{00}^{(2)},
\qquad P_{11}\equiv O_{11,ij}^{(1)} O_{11,ij}^{(2)},
\qquad
P_{02}\equiv \sum_{ij} O_{02,ij}^{(1)} O_{02,ij}^{(2)}, 
\end{equation}
and, using the correspondence (\ref{corre}), 
\begin{eqnarray}
&& O_{00}^{(a)}= \sum_{ei} \phi_{ei}^{(a)} \phi_{ei}^{(a)},
\qquad O_{11,ij}^{(a)} = \phi_{1i}^{(a)} \phi_{2j}^{(a)}
- \phi_{1j}^{(a)} \phi_{2i}^{(a)},
\label{quop} \\
&& O_{02,ij}^{(a)} = \sum_e  \phi_{ei}^{(a)} \phi_{ej}^{(a)}
- \frac{1}{N} \delta_{ij} \sum_{ek} \phi_{ek}^{(a)} \phi_{ek}^{(a)} . 
\nonumber
\end{eqnarray}
The quadratic operator $O_{ml}^{(a)}$ transform 
as a spin-$m$ and a spin-$l$ representation with respect to the ${\rm O}(2)$ and 
${\rm O}(N)$ groups, respectively.  Since $P_{00}$, $P_{11}$, and 
$P_{02}$ belong to different irreducible representations,
they do not mix under RG transformations at the decoupled
${\rm O}(2)\otimes{\rm O}(N)$ FPs.  Their RG dimensions $Y_{ml}$ can be derived
from the RG dimensions $y_{ml}$ of the quadratic operators
$O_{ml}^{(a)}$ at the ${\rm O}(2)\otimes{\rm O}(N)$ FP, using the relation
\begin{equation}
Y_{ml} = 2 y_{ml} - 3.
\label{formyi}
\end{equation}
The quadratic term
$O_{00}^{(a)}$  corresponds to the energy operator and thus $y_{00} = 1/\nu$ and 
$Y_{00}= \alpha/\nu$, where $\alpha$ and $\nu$ are the
specific-heat and correlation-length critical exponents of the
given ${\rm O}(2)\otimes{\rm O}(N)$ FP. The RG dimensions $y_{11}$ and $y_{02}$
were computed in Ref.~\CITE{CPV-05} (there, they are named $y_1$ and 
$y_3$ respectively).

At the chiral FP (1a) we obtain:\cite{foot1}
\[
\begin{array}{llll}
Y_{00} = 0.3(3)\;\;  & Y_{11} = 1.6(3)\;\; & Y_{02} = 0.04(8) & 
   \qquad\qquad\hbox{for }\; N = 3; \\
Y_{00} = 0.2(3)\;\;  & Y_{11} = 1.9(4)\;\; & Y_{02} = -0.4(2) & 
    \qquad\qquad \hbox{for }\; N = 2. 
\end{array}
\]
At the collinear FP (1b) we obtain:\cite{foot1}
\[
\begin{array}{llll}
Y_{00} = 0.3(2)  & Y_{11} = -0.6(2) & Y_{02} = 1.0(3) & 
   \qquad\qquad\hbox{for }\; N = 3; \\
Y_{00} = -0.2182(8)\;\;  & Y_{11} = -0.022(8)\;\; & Y_{02} = 0.9240(11) & 
    \qquad\qquad \hbox{for }\; N = 2. 
\end{array}
\]
These results show that the decoupled ${\rm O}(2)\otimes{\rm O}(N)$ FPs
are unstable in the complete theory (\ref{lgwh}) for both $N=3,2$.

It is also interesting to discuss submodels (a) and (b) mentioned in 
Sec.~\ref{submodels}. In model (a) one should only consider $P_{00}$. 
The numerical results apparently indicate that the FPs are 
always unstable (but, with the present errors, we cannot really exclude the 
opposite possibility), except in one case. For $N=2$,
the collinear FP is stable. In model (b) one should consider $P_{00}$
and $P_{02}$. For $N=2,3$, all FPs are unstable.

\subsection{Stability of the ${\rm O}(4)\otimes{\rm O}(N)$ FPs}
\label{stabo4on}

Here we investigate the stability of FPs (2a) (it does not exist for 
$N=3$) and (2b). For this purpose we must compute the 
RG dimensions of the perturbations of the ${\rm O(4)}\otimes{\rm O(}N{\rm )}$
model appearing in the complete theory. This is done in 
App.~\ref{appo4on}. There are two relevant operators with RG dimensions 
$Y_1$ and $Y_2$. The corresponding perturbative series are reported in 
App.~\ref{appo4on}.  They are analyzed using the conformal
mapping method.\cite{ZJbook,CPV-00}
The errors we will report takes into account the variation of the estimates 
when changing the
resummation parameters $b,\alpha$ defined in Ref.~\CITE{CPV-00}
---we use $b=3,\ldots,18$ and
$\alpha=0,\ldots,4$---and the uncertainty of the FP coordinates.

\begin{figure*}[tb]
\centerline{\psfig{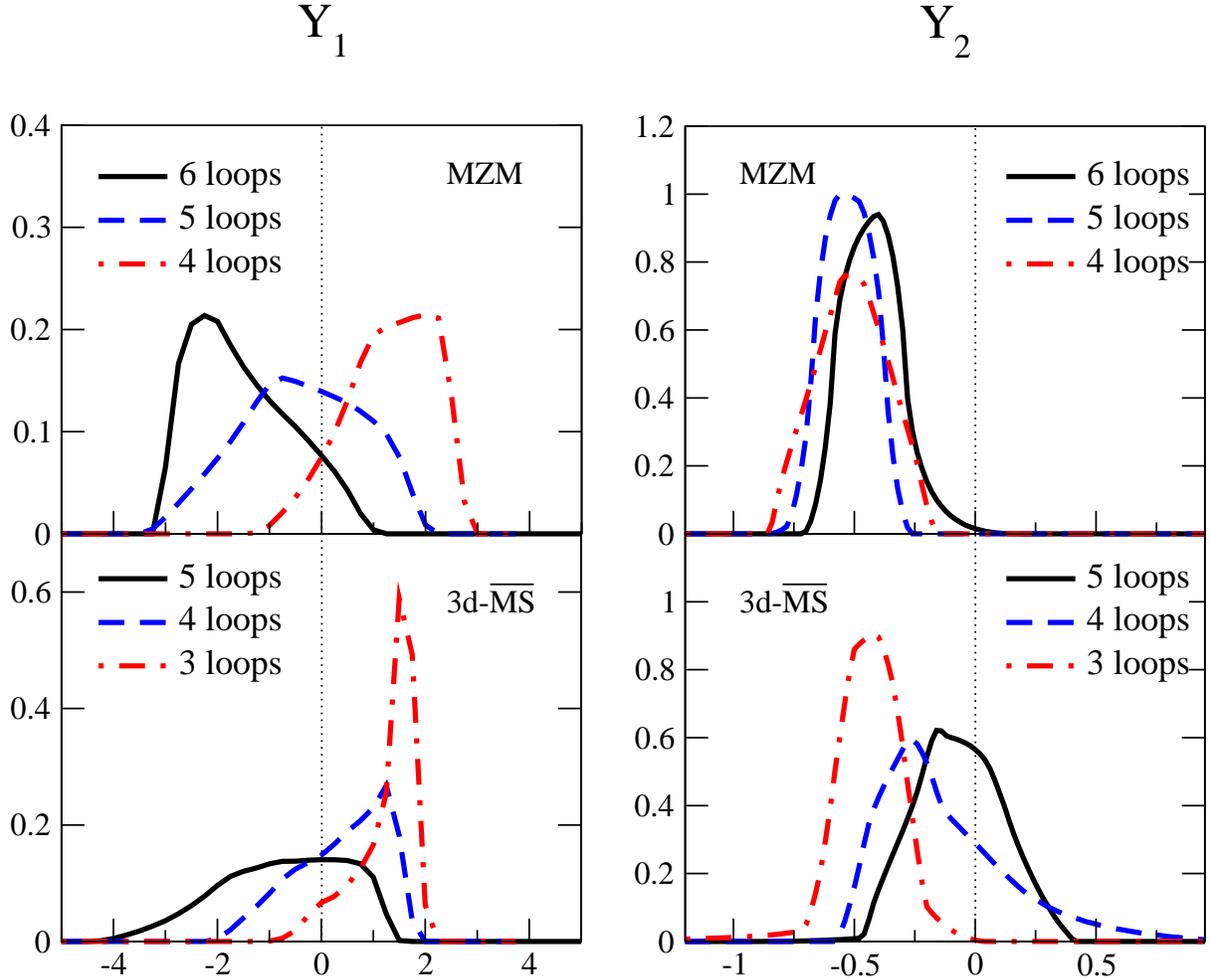}}
\vspace{2mm}
\caption{Distribution of the results for $Y_1$ (left) and $Y_2$ (right)
obtained by 
varying the resummation parameters $\alpha$ and $b$ as a function of the 
number of loops in the MZM and $3d$-$\overline{\rm MS}$ schemes. 
Here $N=3$.}
\label{fig:distribution}
\end{figure*}

The analysis of the six-loop series in the MZM scheme and 
of the five-loop 3$d$-$\overline{\rm MS}$ series gives
the following results at the collinear FP (2b):\cite{foot2}
\begin{equation}
\begin{array}{lll}
Y_1=-0.4(4) \;\; & \quad Y_2=-0.95(7)   & \quad {\rm for} \quad N=2\;\; 
         {\rm (MZM)},\\
Y_1=-0.6(9) \;\; & \quad Y_2=-1.2(1.0)  & \quad {\rm for} \quad N=2\; \;
        \hbox{($3d$-$\overline{\rm MS}$)}, \\
Y_1=-1.5(1.2) \;\; & \quad Y_2=-0.42(10)  & \quad {\rm for} \quad N=3\;\; 
         {\rm (MZM)}, \\
Y_1=-0.8(1.5) \;\; & \quad Y_2=-0.1(2)  & \quad {\rm for} \quad N=3\; \;
        \hbox{($3d$-$\overline{\rm MS}$)}. \\
\end{array}
\end{equation}
For $N=3$ the MZM and $3d$-$\overline{\rm MS}$ results are consistent and
apparently indicate that $Y_1$ and $Y_2$ are negative, though with somewhat
large errors. A better understanding of the relevance of the two operators can
be obtained from Fig.~\ref{fig:distribution}, where we give the distributions
of the estimates of $Y_1$ and $Y_2$ obtained by varying the parameters
$\alpha$ and $b$. For $Y_1$ low-order calculations predict $Y_1 > 0$. However,
as the number of loops increases, $Y_1$ decreases. The six-loop MZM results
indicate that $Y_1 < 0$, a result that is also supported by the trend observed
in the $3d$-$\overline{\rm MS}$ results. As for $Y_2$, the MZM results clearly
indicate $Y_2 < 0$. However, this is not fully confirmed by the
$3d$-$\overline{\rm MS}$ results. Though they give $Y_2 < 0$, there is a trend
towards larger values of $Y_2$. Overall, these results support the
stability of the collinear FP (2b) in the complete theory for $N = 3$.

Similar conclusions hold for $N=2$.
For completeness, we report here the corresponding critical 
exponents:\cite{footnoteQCD}
\begin{equation}
\begin{array}{lll}
\nu=0.71(7) \;\; & \quad \eta=0.12(1)   & \quad {\rm for} \quad N=2\;\; 
         {\rm (MZM)};\\
\nu=0.76(10) \;\; & \quad \eta=0.11(6)  & \quad {\rm for} \quad N=2\; \;
        \hbox{($3d$-$\overline{\rm MS}$)}; \\
\nu=0.89(16) \;\; & \quad \eta=0.18(3)  & \quad {\rm for} \quad N=3\;\; 
         {\rm (MZM)}; \\
\nu=0.88(22) \;\; & \quad \eta=0.10(10)  & \quad {\rm for} \quad N=3\; \; 
\hbox{($3d$-$\overline{\rm MS}$)}\; . \\
\end{array}
\end{equation}
For $N=2$ we also study the stability of the chiral FP (2a).
Using the results of Refs.~\CITE{CPPV-04,CPS-02} for the FP
we have:\cite{footchirale}
\begin{equation}
\begin{array}{lll}
Y_1= -0.03(7) \;\; & \quad Y_2= 0.9(2)     & \quad {\rm (MZM)};\\
Y_1= -0.2(3)  \;\; & \quad Y_2= 0.73(15)   
          & \quad \hbox{($3d$-$\overline{\rm MS}$)}. \\
\end{array}
\end{equation}
The chiral FP is clearly unstable. 

Finally, note that the same discussion also applies to submodel (b), 
since the stability of the FP depends on the same operators with RG 
dimensions $Y_1$ and $Y_2$. For submodel (a) one should only consider 
$Y_1$. In this case also the chiral FP (2b) might be stable.

\subsection{Stability of the $mn$ FP for $N=2$}
\label{stabmn}

Here we investigate the stability of FP (3b) for $N=2$ (it does not exist for 
$N=3$). For this purpose we must compute the 
RG dimensions of the perturbations of the $mn$ FP
appearing in the complete theory. This is done in 
App.~\ref{appmn}. There are two relevant operators with RG dimensions 
$Y_1$ and $Y_2$.
The analysis of the perturbative series in the MZM scheme gives
\begin{eqnarray}
Y_1=-4.0(2.6), \quad Y_2=-0.6(2).
\end{eqnarray}
The results in the $3d$-${\overline{\rm MS}}$ scheme are very imprecise,
although negative values for $Y_1$ and $Y_2$ seem to be favored. 
There results indicate, although with limited confidence, that 
the $mn$ FP present for $N=2$ may be stable in the complete theory.

\section{Conclusions}
\label{conclusions}

In this paper we have studied the quantum phase transition that occurs
in two-dimensional systems that exhibit an ordered phase with SDW
order. The effective Hamiltonian of the relevant critical modes
$\Phi_{ai}$ is given in Eq.~(\ref{lgwhor}).  A detailed mean-field
analysis shows that in some parameter region Hamiltonian
(\ref{lgwhor}) has a continuous transition separating a spin
disordered phase from an ordered phase characterized by two
collinearly polarized SDWs. There are two different possibilities for
the the axes of these SDWs: either $\Phi_1 \| \Phi_2$ or
$\Phi_1 \bot \Phi_2$. We have then investigated the role of
fluctuations in a simplified model in which the two SDWs have the same
velocity. For this purpose we have generated six-loop perturbative
series in the MZM scheme and five-loop series in dimensional
regularization with minimal subtraction.  Close to four dimensions, an
analytic $\epsilon$-expansion calculation shows no presence of stable
FPs. However, past experience indicates that a FP may exist in three
dimensions and be absent for $\epsilon \ll 1$.  Therefore, we have
considered two strictly three-dimensional schemes.  We have analyzed
the stability of some FPs that belong to known submodels. The analysis
of the perturbative series supports the stability of the ${\rm
O(4)}\otimes{\rm O(3)}$ collinear FP.  The analyses of the MZM and
$3d$-$\overline{\rm MS}$ expansions do not provide sufficiently stable
results for the RG flow in the full theory, i.e. in the general space
of its five quartic couplings. In particular, they do not allow us
to draw any definite conclusion on the existence of other stable FPs.
In any case, even without the analysis of the full flow, simple
considerations (reported in App.~\ref{AppA}) show that systems with
collinear SDWs with the same axis (the mean-field solution (3)
reported in Sec.~\ref{sec2}) are in the attraction domain of the ${\rm
O(4)}\otimes{\rm O(3)}$ collinear FP.

It should be remarked that our RG analysis is only valid for $v_1 = v_2$.
In order to extend the results to the generic case $v_1\not=v_2$ one should 
also consider the operator
\begin{equation}
O_v = |\partial_x \Phi_1|^2 - |\partial_y \Phi_1|^2 
    - |\partial_x \Phi_2|^2 + |\partial_y \Phi_2|^2
\end{equation}
and determine its RG dimension $y_v$ at the 
 ${\rm O(4)}\otimes{\rm O(3)}$ collinear FP. If $y_v < 0$ the previous 
conclusions are unchanged. On the other hand, if $y_v > 0$ the 
${\rm O(4)}\otimes{\rm O(3)}$ collinear FP is unstable with respect to 
the perturbation $O_v$. In this case the transition may be of first
order or continuous depending on the existence and attraction domain 
of a stable FP with $v_1\not= v_2$. 
Note that, from a practical point of view, our results are of interest 
even if $y_v > 0$. Indeed, one expects the SDW velocities $v_1$ and $v_2$
to be close in magnitude, of the order of the spin-wave velocity of the 
N\'eel state of the undoped insulator.\cite{ZDS-02} Therefore, the RG flow always
starts very close to the stable FP of the theory with $v_1 = v_2$, and thus
the critical behavior is controlled by this FP  except in a narrow interval
around the critical doping.

Experiments indicate that the SDW-SC--to--SC transition is continuous
and is associated with collinear SDWs. It is thus natural to conjecture
that its critical behavior is controlled by the 
${\rm O(4)}\otimes{\rm O(3)}$ collinear FP, since this FP is stable 
in model (\ref{lgwh}) and its basin of attraction includes 
systems with collinear SDWs.  The corresponding 
critical exponents are then predicted to be $\nu = 0.9(2)$, $\eta = 0.15(10)$.

\section*{Acknowledgments}

MDP acknowledges that part of this work was done at the University of
Roma Tre as part of his PhD thesis.

\appendix

\section{Ground-state configurations} \label{AppA}

In this Appendix we compute the possible ground-state configurations
of Hamiltonian (\ref{lgwh}), that allow us to identify the possible
symmetry-breaking patterns. We consider translation-invariant
configurations and the space-independent Hamiltonian density
\begin{equation}
H(\Phi_1,\Phi_2) = r (|\Phi_1|^2 + |\Phi_2|^2) + H_4(\Phi_1,\Phi_2),
\end{equation}
where $H_4$ is the part of the Hamiltonian that is quartic in the fields.
Since $H_4\ge 0$ for stability, for $r > 0$ the ground state always 
corresponds to $\Phi_1 = \Phi_2 = 0$. For $r < 0$, $\Phi_1 = \Phi_2 = 0$ is 
a local maximum of $H$ and thus the ground state is nontrivial. 
The value $r = 0$ corresponds to a second-order transition point in 
the mean-field approximation. In order to determine the ground states
for $r < 0$, we will first determine all stationary points of $H$; 
the ground state is the one with the lowest energy. Note that, if
$\overline{\Phi}_1$, $\overline{\Phi}_2$ is a stationary point, then
\begin{equation}
H(\overline{\Phi}_1,\overline{\Phi}_2) = {r\over2} 
   ( |\overline{\Phi}_1|^2 + |\overline{\Phi}_2|^2 ) = 
   - H_4(\overline{\Phi}_1,\overline{\Phi}_2).
\label{Hstat}
\end{equation}
This relation is quite general. Indeed, assume $H$ to be of the form
\begin{equation}
    H = \sum_{ij} r_{ij} \phi_i\phi_j + 
        \sum_{ijkl} g_{ijkl} \phi_i\phi_j\phi_k\phi_l.
\end{equation}
Then
\begin{equation}
H = {1\over 4} \sum_i \phi_i {\partial H\over \partial \phi_i} + 
    {1\over2} \sum_{ij} r_{ij} \phi_i\phi_j = 
    {1\over 2} \sum_i \phi_i {\partial H\over \partial \phi_i} -
    \sum_{ijkl} g_{ijkl} \phi_i\phi_j\phi_k\phi_l.
\end{equation}
On a stationary solution, the derivative vanishes, proving Eq.~(\ref{Hstat}).

The calculation of the ground states also allows us to determine the stability
domain of the Hamiltonian. Indeed, a point in the coupling space does not 
belong to the stability domain if there is a field such that $H_4 < 0$. 
Being $H_4$ homogeneous, 
it is not restrictive to consider only fields such that 
$|\Phi_1|^2 + |\Phi_2|^2 = 1$. Thus, the determination of the minima of 
$H_4$ is equivalent to the determination of the minima of $(H - r)$ where 
$r$ is now interpreted as a Lagrange multiplier. Eq.~(\ref{Hstat}) shows
that $H_4$ can be negative only for $r > 0$. Thus, the stability domain
of $H_4$ is obtained by determining the stationary points of $H$ for
$r$ {\em positive}. 

In order to determine the minima, we can use the symmetry of the Hamiltonian.
Using the O($N$) symmetry we can always write 
\begin{equation}
\hbox{Re}\, \Phi_1 = (a,0,\ldots), \qquad
\hbox{Im}\, \Phi_1 = (b,c,0,\ldots).
\label{Phi1-gauge}
\end{equation}
Then, by using the U(1) symmetry we can also fix $b = 0$. Indeed, we 
first perform an O(2) rotation on the first two components:
\begin{equation}
\psi_1' = \psi_1\cos\theta + \psi_2 \sin\theta \qquad
\psi_2' = -\psi_1\sin\theta + \psi_2 \cos\theta,
\end{equation}
where $\psi$ is either $\hbox{Re}\, \Phi_1$ or $\hbox{Im}\, \Phi_1$. 
Then, we apply a U(1) rotation, $\Phi_1'' = e^{i\alpha} \Phi_1'$. If we 
choose 
\begin{equation}
\tan2\theta = {2bc\over a^2 + b^2 - c^2} \qquad
\tan\alpha = {a\sin\theta\over b\sin\theta - c\cos\theta},
\end{equation}
the transformed field has the form (\ref{Phi1-gauge}) with $b = 0$. 
Once $\Phi_1$
has been fixed we can use O$(N-2)$ and U(1) rotations to write
\begin{equation}
\Phi_2 = (d + ie, g + if, l + ih, im, 0, \ldots).
\end{equation}
If $N \le 3$, one can use U(1) rotations to set $l = 0$.

The analysis of the minima is nontrivial due to the complexity of the 
stationarity equations. We have only consider the case $N \le 3$ that is 
relevant experimentally. Other ground states are present 
for $N \ge 4$.
We found seven relevant minima (only five of them occur for $N = 2$):
\begin{itemize}
\item[1)] $a^2 = - r/u_{12}$, $H = - r^2/(2 u_{12})$.
\item[2)] $a^2 = c^2 = - r/u_{1}$, $H = - r^2/(2 u_{1})$.
\item[3)] $a^2 = d^2 = - r/(u_{12} + w_1 + w_+)$, 
          $H = - r^2/(u_{12} + w_1 + w_+)$.
\item[4)] $a^2 = f^2 = - r/(u_{12} + w_1)$, 
          $H = - r^2/(u_{12} + w_1)$.
\item[5a,b)] $a^2 = c^2 = d^2 = f^2 = - r/(2 u_{1} + 2 w_1 + w_+ \pm w_-)$, 
          $H = - 2 r^2/(2 u_{1} + 2 w_1 + w_+ \pm w_-)$.
\item[6)] [$N\ge 3$] $c^2 = -r (u_1 - w_1)/\Delta_6$, 
     $d^2 = h^2 = -r (u_{12} - w_1)/(2\Delta_6)$,
     $H = -r^2 (u_1 + u_{12} - 2 w_1)/(2\Delta_6)$,
     $\Delta_6 = u_1 u_{12} - w_1^2$.
\item[7)] [$N\ge 3$, $w_\pm \not= 0$] 
$a^2 = d^2 \not= 0$, $c,f,h\not=0$, $ad/(cf) = w_+/w_-$,
$a^2 + c^2 = d^2 + f^2 + h^2$, with energy
$H = - r^2 [u_2 (w_2 + w_3) + w_2 w_3]/\Delta_7$,
with $\Delta_7 = (u_{12} + w_1) w_2 w_3 +  w_+ u_2 (u_1 + w_1)$.
Alternatively, if we define the four vectors 
$t_1 = {\rm Re}\, \Phi_1$, 
$t_2 = {\rm Im}\, \Phi_1$,
$t_3 = {\rm Re}\, \Phi_2$, 
$t_4 = {\rm Im}\, \Phi_2$,
and $t_{ij} = t_i\cdot t_j$,
the solution can be characterized more geometrically as follows:
$t_{11} = t_{22} = t_{33} = t_{44} = H/(2 r)$, 
$t_{12} = t_{34}$, $t_{13} = t_{24}$, and $t_{14} = t_{23}$, with
\begin{eqnarray}
t_{12}^2 = r^2 w_2^2 w_3^2/\Delta_7^2 \qquad
t_{13}^2 = r^2 u_2^2 w_2^2/\Delta_7^2 \qquad
t_{14}^2 = r^2 u_2^2 w_3^2/\Delta_7^2.
\end{eqnarray}
\end{itemize}
Whenever a component is not explicitly written, it vanishes. Moreover,
we defined $u_{12} \equiv  u_1 + u_2$, $w_+ \equiv  w_2 + w_3$,
$w_- \equiv  w_2 - w_3$ and we simplified the notation
writing $u_1$ instead of $u_{1,0}$, etc. Beside the seven solutions reported above, 
for $w_- \not = 0$ we also
found stationary points with $e = g = h = 0$ and 
\begin{equation}
{df\over ac} = {- 4 u_2^2 - w_-^2 + w_+^2 \pm
      \sqrt{\vphantom{P^2} (4 u_2^2 + w_-^2 - w_+^2)^2 - 16 u_2^2 w_-^2} \over 
      4 u_2 w_-}.
\end{equation}
A numerical analysis indicates that they are never 
absolute minima of the Hamiltonian and thus they are  never relevant
for the ground-state calculation. For this reason, these solutions have not 
been included above. The computation of all stationary points is quite 
straightforward, except for solution 7. We shall now briefly sketch 
how it is derived. Assume that $e = g = 0$ and $a,c,d,f,h\not=0$ and 
define $E_a = (1/a) {\partial H/\partial a}$, etc. 
Then,
\begin{equation}
 {f\over c} (E_f - E_h) = - a d w_- + c f w_+ = 0.
\label{eq1:AppA}
\end{equation}
Using this relation, we can rewrite $E_a$, $E_c$, $E_d$, and $E_f$ as 
linear equations in $a^2$, $c^2$, $d^2$, $f^2$, and $h^2$. 
Considering also  $a^2 d^2 w_-^2 = c^2 f^2 w_+^2$ that follows from
Eq.~(\ref{eq1:AppA}), we obtain a system of equations that allows us
to determine all components.

Given the list of solutions, we can determine the stability domain 
of the Hamiltonian. Using solutions 1-5, we obtain the necessary conditions 
\begin{eqnarray}
&& u_1 > 0, \qquad\qquad 
   u_{12} > 0, \qquad\qquad
   u_{12} + w_1 > 0, \qquad\qquad
\nonumber \\
&& u_{12} + w_1 + w_+ > 0, \qquad\qquad
   u_{1} + w_1 + {1\over2} (w_+ \pm w_-) > 0.
\label{stability-cond-n2}
\end{eqnarray}
These conditions are sufficient for $N=2$. For $N\ge 3$ we must also consider
solutions 6 and 7. Solution 6 gives the necessary condition
\begin{equation}
  w_1 > - \sqrt{u_1 u_{12}} .
\label{stability-cond-n3}
\end{equation}
Numerically, we find that solution 7 is also relevant for stability,
although we have not been able to write down an easy condition.

For cuprates the relevant solutions are 3 and 4. In view of the possibility that
the ${\rm O(4)}\otimes {\rm O(3)}$ 
 FP is stable it is important to understand to which 
ground state of the ${\rm O(4)}\otimes {\rm O(3)}$ Hamiltonian 
(\ref{Hch}) they correspond. For generic $n$ and $m$, $m \ge n$, 
model (\ref{Hch}) is stable for 
$g_{1,0} > 0$ and $n g_{1,0} - (n-1) g_{2,0} > 0$ and has two ground states
depending on the sign of $g_{2,0}$: for $g_{2,0}>0$ the ground state 
is chiral, while for $g_{2,0}<0$ the ground state is 
collinear.\cite{Kawamura-98} The corresponding energies are: 
\begin{equation}
\begin{array}{ll}
 H = - \displaystyle{3 n r^2 \over 2 [ n g_{1,0} - (n-1) g_{2,0}]} & \qquad\qquad 
 \hbox{(chiral)}; 
\\
 \vphantom{a} & \\
 H = - \displaystyle{3 r^2 \over 2 g_{1,0}} & \qquad\qquad 
 \hbox{(collinear)}.
\end{array}
\end{equation}
Using $u_{12} = g_{1,0}/3$, $u_2 = g_{2,0}/6$, $w_1 = (g_{1,0} - g_{2,0})/3$, 
$w_+ = g_{2,0}/3$, $w_- = 0$, we immediately see that for $m=4$ and $n=2$ and $n=3$ 
solutions 1 and 3 correspond to the collinear case.
For $n=3$ 
solutions 6 and 7 correspond to the chiral case, while 
solutions 2, 4, and 5 correspond to a stationary state that is never
a ground state in the chiral theory. For $n=2$ instead, solutions 2, 4, and 5
are those corresponding to the chiral case. This result is relevant to identify
the attraction domain of the 
${\rm O(4)}\otimes {\rm O(3)}$ collinear FP in the full theory.
Indeed, it shows that the attraction domain of this FP 
includes systems whose ground state is given by solutions 1 and 3 (and
therefore two collinearly polarized SDWs).
Nothing can be said on the other solutions: in this case an analysis of 
the RG flow of the full theory is needed.

\section{Renormalization-group dimensions of the perturbations at the 
O(4)$\otimes$O($N$) fixed points} \label{appo4on}

We need to classify the operators that break 
\begin{equation}
{\rm O}(4)\otimes{\rm O}(N) \to ({\rm U}(1)\oplus{\rm U}(1))\otimes{\rm O}(N) \cong
            ({\rm SO}(2)\oplus{\rm SO}(2))\otimes{\rm O}(N)\; .
\end{equation}
This is essentially discussed in Ref.~\CITE{CPV-05}. 
There are, however, two differences:
first, we have only SO(2) symmetry, instead of O(2) symmetry; 
second, there is an additional exchange symmetry that forbids the appearance
of spin-2 operators. In the notations of Ref.~\CITE{CPV-05} ($M$ and $N$ of 
Ref.~\CITE{CPV-05} correspond to $N$ and 4 respectively) we define
\begin{eqnarray}
&&
P_1\equiv {\cal O}^{(4,4)}_{1133}+{\cal O}^{(4,4)}_{1144}+
{\cal O}^{(4,4)}_{2233}+{\cal O}^{(4,4)}_{2244},
\label{perto4on} \\ 
&&
P_2 \equiv  {\cal O}^{(4,r)}_{1313}+{\cal O}^{(4,r)}_{1414}+
{\cal O}^{(4,r)}_{2323}+{\cal O}^{(4,r)}_{2424},
\nonumber \\
&&
P_3 \equiv {\cal O}^{(4,r)}_{1234},
\nonumber
\end{eqnarray}
where $\phi_{ei}$ is the real field defined in Eq.~(\ref{o4oncorr}). Note that
$P_3$ would be forbidden if we had O(2) invariance instead of SO(2) invariance.
Moreover, $P_2$ and $P_3$ correspond to different components of the 
same operator, so that they have the same RG dimension.

Hamiltonian (\ref{lgwh}) can then be written as 
\begin{eqnarray}
\mathcal{H}&=&\int{d^d x}\, 
   \sum_{a}\frac{1}{2}\left[ (\nabla \phi_{a})^2+ \phi_{a}^2 \right] +
   t_1 ( \sum_a \phi_{a}^2 )^2
 + t_2\sum_{a,b} \left[ (\phi_a \cdot \phi_b)^2 -\phi_a^2 \phi_b^2 \right]
\nonumber \\
&& + t_3 P_1 + t_4 P_2 + t_5 P_3,
\end{eqnarray}
where 
\begin{eqnarray}
&&
t_1={1\over 24} (2 u_1 + 2 u_2 + w_1 + w_2 + w_3)
\nonumber  \\
&&
t_2={1\over36} (u_1 + 4 u_2 - w_1 + 2 w_2 + 2 w_3)
\nonumber \\
&&
t_3={1\over 12} (-u_1-u_2+w_1+w_2+w_3)
\nonumber \\
&&
t_4={1\over 12} (-2u_1+4u_2+2 w_1-w_2-w_3)
\nonumber \\
&&
t_5={1\over2} (w_3-w_2)\ .
\label{cho4on} 
\end{eqnarray}
Since all operators are irreducible with respect to ${\rm O}(4)\otimes{\rm O}(N)$
transformations, if the couplings belong to the ${\rm O}(4)\otimes{\rm O}(N)$ theory,
the stability matrix defined with respect to the couplings $t_i$ has the 
form
\begin{equation}
\Omega=\left( \begin{array}{ccccc}
\Omega_{11} & \Omega_{21} & 0 & 0 & 0 \\
\Omega_{21} & \Omega_{22} & 0 & 0 & 0 \\
0 & 0 & \Omega_{1} & 0 & 0 \\
0 & 0 & 0 & \Omega_{2} & 0 \\
0 & 0 & 0 & 0 & \Omega_{2} \\
\end{array} \right) 
\end{equation}
Here $Y_1 = - \Omega_1$ is the RG dimension of ${\cal O}^{(4,4)}_{abcd}$ and 
$Y_2 = - \Omega_2$ is the RG dimension of ${\cal O}^{(4,r)}$.

In the MZM scheme for $N=2$ we find:
\begin{eqnarray}
\Omega_{1}&=&1-(0.238732\,u_1 + 0.31831\,u_2)+\nonumber\\
&&+(0.0324838\,u_1^2 + 0.0614494\,u_1\,u_2 +
0.0342428\,u_2^2)+\nonumber\\ &&-(0.00570145\,u_1^3 +
0.016556\,u_1^2\,u_2 + 0.0218393\,u_1\,u_2^2 +
0.00451785\,u_2^3)+\nonumber\\ &&+(0.00156398\,u_1^4 +
0.0052904\,u_1^3\,u_2 + 0.0097641\,u_1^2\,u_2^2 +
0.00663506\,u_1\,u_2^3 + \nonumber\\ && +
0.000911563\,u_2^4)-(0.00045218\,u_1^5 + 0.00188296\,u_1^4\,u_2 +
0.00448727\,u_1^3\,u_2^2 \nonumber\\ &&+ 0.00487778\,u_1^2\,u_2^3 +
0.00209129\,u_1\,u_2^4 + 0.000250855\,u_2^5)+ (0.000158666\,u_1^6
+\nonumber\\ &&+0.000747711\,u_1^5\,u_2 + 0.00212573\,u_1^4\,u_2^2 +
0.00311938\,u_1^3\,u_2^3 + 0.00230887\,u_1^2\,u_2^4 +\nonumber\\ &&+
0.000759722\,u_1\,u_2^5 + 0.0000778366\,u_2^6),
\end{eqnarray}
\begin{eqnarray}
\Omega_{2}&=&1-(0.238732\,u_1 - 0.159155\,u_2)+\nonumber\\
&&+(0.0324838\,u_1^2 - 0.0145415\,u_1\,u_2 -
0.0164178\,u_2^2)+\nonumber\\ &&-(0.00570145\,u_1^3 -
0.00426028\,u_1^2\,u_2 - 0.00222856\,u_1\,u_2^2 +
0.00294388\,u_2^3)+\nonumber\\ &&+(0.00156398\,u_1^4 -
0.000683968\,u_1^3\,u_2 + 9.94721\cdot {10}^{-6}\,u_1^2\,u_2^2 +
0.000950042\,u_1\,u_2^3 +\nonumber\\ &&-
0.000152361\,u_2^4)-(0.00045218\,u_1^5 - 0.000214037\,u_1^4\,u_2 +
0.00011481\,u_1^3\,u_2^2 + \nonumber\\ &&+ 0.000428198\,u_1^2\,u_2^3 -
0.0000869867\,u_1\,u_2^4 - 0.0000326009\,u_2^5)+(0.000158666\,u_1^6
+\nonumber\\ &&-0.000038529\,u_1^5\,u_2 + 0.000128979\,u_1^4\,u_2^2 +
0.000233286\,u_1^3\,u_2^3 - 0.0000225715\,u_1^2\,u_2^4 +\nonumber\\
&&- 0.0000164528\,u_1\,u_2^5 + 8.9658\cdot{10}^{-6}\,u_2^6).
\end{eqnarray}
For $N=3$ we obtain:
\begin{eqnarray}
\Omega_{1}&=&1-(0.238732\,u_1 + 0.397887\, u_2)+\nonumber\\
&&+(0.0378782\,u_1^2 + 0.0687202\,u_1\,u_2 +
0.0413963\,u_2^2)+\nonumber\\ &&-(0.00623158\,u_1^3 +
0.0193945\,u_1^2\,u_2 + 0.0290399\,u_1\,u_2^2 +
0.00434652\,u_2^3)+\nonumber\\ &&+(0.0020362\,u_1^4 +
0.00594275\,u_1^3\,u_2 + 0.0126485\,u_1^2\,u_2^2 +
0.00940724\,u_1\,u_2^3 + \nonumber\\
&&+0.000356974\,u_2^4)-(0.000572885\,u_1^5 + 0.00220348\,u_1^4\,u_2 +
0.00583754\,u_1^3\,u_2^2 \nonumber\\ &&+ 0.0069323\,u_1^2\,u_2^3 +
0.00263869\,u_1\,u_2^4 + 0.000106569\,u_2^5) +(0.000227195\,u_1^6
+\nonumber\\ &&+ 0.000881969\,u_1^5\,u_2 + 0.00277192\,u_1^4\,u_2^2 +
0.00439833\,u_1^3\,u_2^3 + 0.00335526\,u_1^2\,u_2^4 +\nonumber\\ &&+
0.000841582\,u_1\,u_2^5 + 0.0000352387\,u_2^6),
\end{eqnarray}
\begin{eqnarray}
\Omega_{2}&=&1-(0.238732\,u_1 - 0.0795775\,u_2)+\nonumber\\
&&+(0.0378782\,u_1^2 - 0.00727073\,u_1\,u_2 -
0.0314283\,u_2^2)+\nonumber\\ &&-(0.00623158\,u_1^3 -
0.00255553\,u_1^2\,u_2 - 0.00424767\,u_1\,u_2^2 +
0.00445854\,u_2^3)+\nonumber\\ &&+(0.0020362\,u_1^4 -
0.000329962\,u_1^3\,u_2 - 0.000353201\,u_1^2\,u_2^2 +
0.00115746\,u_1\,u_2^3 +\nonumber\\ &&-
0.000132244\,u_2^4)-(0.000572885\,u_1^5 - 0.000151507\,u_1^4\,u_2 +
0.000153088\,u_1^3\,u_2^2 + \nonumber\\ &&+ 0.000587068\,u_1^2\,u_2^3
- 0.0000150878\,u_1\,u_2^4 - 0.0000901012\,u_2^5)+(0.000227195\,u_1^6
+ \nonumber\\ &&- 0.0000200211\,u_1^5\,u_2 + 0.000205186\,u_1^4\,u_2^2
+ 0.000321211\,u_1^3\,u_2^3 + 0.000031431\,u_1^2\,u_2^4 +\nonumber\\
&&- 0.0000663864\,u_1\,u_2^5 + 0.0000193321\,u_2^6).
\end{eqnarray}

In the $3d$-$\overline{\rm MS}$ we find for $N=2$:
\begin{eqnarray}
\Omega_{1}&=&1-(6\,u_1 + 8\,u_2)+\nonumber\\ &&+(30.5\,u_1^2 +
58\,u_1\,u_2 + 32\,u_2^2)+\nonumber\\ &&-(327.297\,u_1^3 +
928.74\,u_1^2\,u_2 + 1185.79\,u_1\,u_2^2 + 258.397\,u_2^3)+\nonumber\\
&&+(5835.31\,u_1^4 + 20132.4\,u_1^3\,u_2 + 35648.3\,u_1^2\,u_2^2 +
23296.2\,u_1\,u_2^3 + \nonumber\\ && + 4377.86\,u_2^4 )-(
123668\,u_1^5 + 506531\,u_1^4\,u_2 + 1.1389\cdot{10}^6\,u_1^3\,u_2^2 +
\nonumber\\ && +1.1852\cdot{10}^6\,u_1^2\,u_2^3 + 552355\,u_1\,u_2^4 +
85949\,u_2^5),
\end{eqnarray}
\begin{eqnarray}
\Omega_2&=&1-(6\,u_1 - 4\,u_2)+\nonumber\\ &&+(30.5\,u_1^2 -
14\,u_1\,u_2 - 16\,u_2^2)+\nonumber\\ &&-(327.297\,u_1^3 -
219.798\,u_1^2\,u_2 - 163.596\,u_1\,u_2^2 +
103.301\,u_2^3)+\nonumber\\ &&+(5835.31\,u_1^4 - 3001.7\,u_1^3\,u_2 -
2298.63\,u_1^2\,u_2^2 + 1415.85\,u_1\,u_2^3 + \nonumber\\ && +
198.484\,u_2^4)-( 123668\,u_1^5 - 48887.2\,u_1^4\,u_2 -
24466.2\,u_1^3\,u_2^2 + \nonumber\\ &&+ 34690.3\,u_1^2\,u_2^3 -
12396.4\,u_1\,u_2^4 - 4256.31\,u_2^5).
\end{eqnarray}
and finally for $N=3$:
\begin{eqnarray}
\Omega_{1}&=&1-(6\,u_1 + 10\,u_2)+\nonumber\\ &&+(35.5\,u_1^2 +
65\,u_1\,u_2 + 38.5\,u_2^2)+\nonumber\\ &&-(369.646\,u_1^3 +
1082.64\,u_1^2\,u_2 + 1548.49\,u_1\,u_2^2 +
251.598\,u_2^3)+\nonumber\\ &&+( 7381.82\,u_1^4 + 23673.\,u_1^3\,u_2 +
45677.7\,u_1^2\,u_2^2 + 31394.4\,u_1\,u_2^3 + \nonumber\\ && +
3819.3\,u_2^4)-(169602\,u_1^5 + 615152\,u_1^4\,u_2 +
1.48093\cdot{10}^6\,u_1^3\,u_2^2 + \nonumber\\ &&+
1.61998\cdot{10}^6\,u_1^2\,u_2^3 + 719990\,u_1\,u_2^4 + 76742\,u_2^5 ),
\end{eqnarray}
\begin{eqnarray}
\Omega_{2}&=&1-(6\,u_1 - 2\,u_2)+\nonumber\\ &&+(35.5\,u_1^2 -
7\,u_1\,u_2 - 30.5\,u_2^2)+\nonumber\\ &&-(369.646\,u_1^3 -
130.399\,u_1^2\,u_2 - 306.317\,u_1\,u_2^2 + 164.3\,u_2^3)+\nonumber\\
&&+( 7381.82\,u_1^4 - 1861.44\,u_1^3\,u_2 - 4739.46\,u_1^2\,u_2^2 +
2138.26\,u_1\,u_2^3 + \nonumber\\ && + 581.765\,u_2^4)-(
169602\,u_1^5 - 33233.6\,u_1^4\,u_2 - 55928.8\,u_1^3\,u_2^2 +
\nonumber\\ &&+ 53320.3\,u_1^2\,u_2^3 - 13607.5\,u_1\,u_2^4 -
6933.24\,u_2^5).
\end{eqnarray}

\section{Renormalization-group dimensions of the perturbations at the 
${mn}$ fixed points} \label{appmn}

The analysis of the perturbations at the $mn$ FP is quite simple. 
In our case $m = 2 N$, $n = 2$ and the relevant symmetry group is 
O($2N$), which is broken by the terms proportional to 
$u_2$, $w_2$, and $w_3$. If $\phi_{ai}$ is the field defined in 
Eq.~(\ref{corr-mn}), $a=1,2$, $i=1,\ldots 2N$, we define the following 
spin-2 and spin-4 operators that transform irreducibly under O($2N$):
\begin{eqnarray}
V_{a,i,j}^{(2)} &\equiv& \phi_{ai} \phi_{aj} - {1\over 2N} \delta_{ij} \phi^2_a 
\\
V_{a,i,j,k,l}^{(4)} &\equiv & \phi_{ai} \phi_{aj} \phi_{ak} \phi_{al} 
- {1\over 2(N+2)} \phi^2_a (\delta_{ij} \phi_{ak} \phi_{al} + 
      \hbox{5 perm.}) \nonumber \\
&& + {1\over 4 (N+1)(N+2)} (\phi^2_a)^2 
       (\delta_{ij} \delta_{kl} + \delta_{ik} \delta_{jl} + 
        \delta_{il} \delta_{jk}),
\end{eqnarray}
where $\phi_a^2 \equiv  \sum_i\phi_{ai}^2$. Then, the relevant operators are:
\begin{eqnarray}
P_1 &\equiv & \sum_{a=1}^2\sum_{\alpha\beta=0}^1 \sum_{ij=1}^N
   V_{a,i+\alpha N,i+\beta N,j+\alpha N,j+\beta N}^{(4)},
\\
P_2 &\equiv & \sum_{\alpha\beta=0}^1 \sum_{ij=1}^N
   V_{1,i+\alpha N,j+\alpha N}^{(2)}  
   V_{2,i+\beta N,j+\beta N}^{(2)}  ,
\\
P_3 &\equiv & \sum_{\alpha\beta\gamma\delta=0}^1 \sum_{ij=1}^N
   \epsilon_{\alpha\beta} \epsilon_{\gamma\delta}
   V_{1,i+\alpha N,j+\beta N}^{(2)}  
   V_{2,i+\gamma N,j+\delta N}^{(2)}  ,
\end{eqnarray}
where $\epsilon_{01} = -\epsilon_{10} = 1$ and 
$\epsilon_{00} = \epsilon_{11} = 0$.
These operators give rise to different breakings of O($2 N$):
\begin{eqnarray}
{\rm O}(2 N) &\stackrel{P_1}{\longrightarrow} &
  [{\rm O}(N)\otimes{\rm O}(2)]\oplus[{\rm O}(N)\otimes{\rm O}(2)]
   \stackrel{P_2}{\longrightarrow}{\rm O}(N)\otimes[{\rm O}(2)\oplus{\rm O}(2)] 
\nonumber  \\
   &\stackrel{P_3}{\longrightarrow}&{\rm O}(N)\otimes{\rm S}[{\rm O}(2)\oplus{\rm O}(2)].
\end{eqnarray}
In terms of $P_1$, $P_2$, and $P_3$
Hamiltonian (\ref{lgwh}) can then be written as 
\begin{eqnarray}
\mathcal{H}&=&\int{d^d x}\, 
   \sum_{a}\frac{1}{2}\left[ (\nabla \phi_{a})^2+ \phi_{a}^2 \right] +
   t_1 \sum_a (\phi_{a}^2)^2
 + t_2 \phi_1^2 \phi_2^2 
\nonumber \\
&& + t_3 P_1 + t_4 P_2 + t_5 P_3,
\end{eqnarray}
where
\begin{equation}
t_1 = {u_1\over2} + {u_2\over N + 1}, \quad
t_2 = w_1 + {1\over N}(w_2 + w_3), \quad
t_3 = u_2, \quad
t_4 = w_2 + w_3, \quad
t_5 = w_3 - w_2. \quad
\end{equation}
Since all operators are irreducible with respect to ${\rm O}(2N)$
transformations, if the couplings belong to the $mn$ theory,
the stability matrix defined with respect to the couplings $t_i$ has the 
form
\begin{equation}
\Omega=\left( \begin{array}{ccccc}
\Omega_{11} & \Omega_{21} & 0 & 0 & 0 \\
\Omega_{21} & \Omega_{22} & 0 & 0 & 0 \\
0 & 0 & \Omega_{1} & 0 & 0 \\
0 & 0 & 0 & \Omega_{2} & 0 \\
0 & 0 & 0 & 0 & \Omega_{2} \\
\end{array} \right) 
\end{equation}
Note that two eigenvalues are degenerate, since $P_2$ and $P_3$ 
are different components of the same irreducible operator 
$V_{1,i,j}^{(2)}V_{2,k,l}^{(2)}$. The corresponding RG dimensions
are $Y_1 = - \Omega_1$, $Y_2 = - \Omega_2$.


\begin{references}

\bibitem{Sachdev-03}
S. Sachdev, Rev. Mod. Phys. {\bf 75}, 913 (2003).

\bibitem{expt}
S. Wakimoto, G. Shirane, Y. Endoh, K. Hirota, S. Ueki,
K. Yamada, R.J.~Birgeneau, M.A.~Kastner, Y.S.~Lee, 
P.M.~Gehring, and S.H.~Lee, Phys. Rev. B {\bf 60}, R769 (1999);
S.~Wakimoto, R.J.~Birgeneau, Y.S.~Lee, and G. Shirane,
Phys. Rev. B {\bf 63}, 172501 (2001).

\bibitem{AMHMK-97}
G. Aeppli, T.E. Mason, S.M. Hayden, H.A. Mook, and J. Kulda,
Science {\bf 278}, 1432 (1997).

\bibitem{ZDS-02}
Y. Zhang, E. Demler, and S. Sachdev,
Phys. Rev. B {\bf 66}, 094501 (2002).

\bibitem{Yee-etal-99}
Y.S.~Lee, R.J.~Birgeneau, M.A.~Kastner, Y.~Endoh,
S.~Wakimoto, K.~Yamada, R.W.~Erwin, S.-H.~Lee, and 
G.~Shirane, 
Phys. Rev. B {\bf 60}, 3643 (1999).

\bibitem{WF-72} 
K.G. Wilson and M.E. Fisher,
Phys.\ Rev.\ Lett.\  {\bf 28}, 240  (1972).

\bibitem{superc}
See, for example, S. Mo, J. Hove, and A. Sudb\o,
Phys. Rev. B {\bf 65}, 104501 (2002), and references
therein.

\bibitem{CPPV-04}
P. Calabrese, P. Parruccini, A. Pelissetto, and E. Vicari, 
Phys. Rev. B {\bf 70}, 174439 (2004).

\bibitem{DPV-04}
M. De Prato, A. Pelissetto, and E. Vicari,
Phys. Rev. B {\bf 70}, 214519 (2004).

\bibitem{SD-89}
R.~Schloms and V.~Dohm,
Nucl.\ Phys.\ B  {\bf 328}, 639 (1989).

\bibitem{Parisi-80}
G. Parisi, 
{\em Field-theoretic approach to second-order phase transitions in 
two- and three-dimensional systems},
Carg\`{e}se Lectures (1973),
J.\ Stat.\ Phys.\  {\bf 23}, 49 (1980).

\bibitem{tHV-72}
G. 't Hooft and M.J.G. Veltman, 
Nucl. Phys. B  {\bf 44}, 189 (1972).

\bibitem{NMB-77}
B.G.~Nickel, D.I.~Meiron, and G.A.~Baker, Jr.,
{\em Compilation of 2-pt and 4-pt graphs for continuum spin models},
Guelph University Report, 1977, unpublished.

\bibitem{KS-01} 
H.~Kleinert and V.~Schulte-Frohlinde, 
{\em Critical Properties of $\phi^4$-Theories}\/
(World Scientific, Singapore, 2001).

\bibitem{ZKE-98}
O. Zachar, S.A.~Kivelson, and V.J.~Emery,
Phys. Rev. B {\bf 57}, 1422 (1998).

\bibitem{PRV-01b}
A.~Pelissetto, P. Rossi, and E.~Vicari,
Nucl. Phys. B  {\bf 607}, 605 (2001).

\bibitem{Kawamura-98}
H. Kawamura,
J. Phys.: Condens. Matter {\bf 10}, 4707 (1998).

\bibitem{CP-04}
P. Calabrese and P. Parruccini, Nucl. Phys. B {\bf 679}, 568 (2004).

\bibitem{PV-rev}
A. Pelissetto and E. Vicari, Phys. Rep. {\bf 368}, 549  (2002).

\bibitem{DMT-03}
B. Delamotte, D. Mouhanna, and M. Tissier,
Phys. Rev. B {\bf 69}, 134413 (2004).

\bibitem{PRV-01}
A.~Pelissetto, P. Rossi, and E.~Vicari,
Phys. Rev. B  {\bf 63}, 140414(R) (2001).

\bibitem{CPS-02}
P. Calabrese, P. Parruccini, and A.I. Sokolov,
Phys. Rev. B {\bf 66}, 180403(R) (2002);
Phys. Rev. B {\bf 68}, 094415 (2003).

\bibitem{CPV-05}
P. Calabrese, A. Pelissetto, and E. Vicari,
Nucl. Phys. B {\bf 709}, 550 (2005).

\bibitem{Parruccini-03}
P. Parruccini, Phys. Rev. B {\bf 68}, 104415 (2003).

\bibitem{foot2}
For $N=2$ the collinear FP is (Ref.~\CITE{CPV-05})
$u_1=6.3(4)$ and $u_2=-8.3(3)$ (MZM),
$u_1=0.33(3)$ and $u_2=-0.30(2)$ (3$d$-$\overline{\rm MS}$).
For $N=3$ the collinear FP is:
$u_1=4.3(5)$ and $u_2=-7.5(6)$ (MZM),
$u_1=0.23(5)$ and $u_2=-0.23(3)$ (3$d$-$\overline{\rm MS}$).

\bibitem{Aharony-76}
A. Aharony,
In {\em Phase Transitions and Critical Phenomena},
Vol. 6, edited by C. Domb and M.S. Green\/ (New York, Academic, 1976).

\bibitem{PV-05}
A. Pelissetto and E. Vicari,
Cond. Matt. Phys. (Ukraine) {\bf 8}, 87 (2005)
[hep-th/0409214]. 

\bibitem{ZJbook}
J. Zinn-Justin,
{\em Quantum Field Theory and Critical Phenomena},
fourth edition\/
(Clarendon Press, Oxford, 2001).

\bibitem{PV-00}
A. Pelissetto and E. Vicari, Phys. Rev. B {\bf 62}, 6393 (2000).

\bibitem{foot1}
At the chiral FP we use (Refs.~\CITE{CPPV-04,CPV-05}):
$\nu = 0.60(8)$, $y_{11} = 2.3(2)$,   $y_{02} = 1.52(6)$  ($N=3$);
$\nu = 0.63(9)$, $y_{11} = 2.45(25)$, $y_{02} = 1.30(15)$ ($N=2$).
At the collinear FP, for $N=3$ we use (Refs.~\CITE{DPV-04,CPV-05}):
$\nu = 0.60(5)$, $y_{11} = 1.20(15)$, $y_{02} = 2.0(2)$.
For $N=2$, we use the mapping with the XY model (Refs.~\CITE{Kawamura-98,CPV-05})
and the results of M. Campostrini, M. Hasenbusch, A. Pelissetto, P. Rossi,
and E. Vicari, Phys. Rev. B {\bf 63},  214503 (2001),
$\nu = 0.67155(27)$, $y_{11} = 1.489(6)$, $y_{02} = 1.9620(8)$.

\bibitem{CPV-00}
J.~M.~Carmona, A. Pelissetto, and E. Vicari,
Phys. Rev. B {\bf 61}, 15136 (2000).

\bibitem{footchirale}
We use $u_1 = -1.1(5)$ and $u_2 = 6.1(4)$ (MZM), 
$u_1 = 0.03(2)$ and $u_2 = 0.215(13)$ (3$d$-$\overline{\rm MS}$).

\bibitem{footnoteQCD}
It is interesting to note that these exponents are also relevant for 
quantum chromodynamics (the theory of strong interactions)
with two quarks at finite temperature if the 
anomaly contribution at the critical point is small 
[F.~Basile, A.~Pelissetto, and E. Vicari, Proceedings of the 
Symposium on Lattice Field Theory 2005, PoS (LAT2005) 199,
hep-lat/0509018, and
J. High Energy Phys. {\bf 02}, 044 (2005);
A.~Butti, A.~Pelissetto, and E. Vicari, 
J. High Energy Phys. {\bf 08}, 029 (2003)].

\end{references}
\end{document}